\documentclass[12pt,preprint]{aastex62}

\usepackage{amssymb}% Include extra symbols: \gtrsim and \lesssim
\usepackage{fancyhdr}
\usepackage{mathptmx} % Times

\usepackage{aas_macros}
\usepackage{natbib}
\usepackage{graphics,graphicx}
\def\apjl{\rm ApJL}
\def\apjs{\rm ApJS}

\def\aap{\rm A\&A}

\def\checkbox{\square\!\!\!\!\!\!\!{\bf X}}

\usepackage{color}
\usepackage{wrapfig}

\usepackage[compact]{titlesec}
\titlespacing{\section}{0pt}{6pt}{4pt}
\titlespacing{\subsection}{0pt}{6pt}{4pt}
\titlespacing{\subsubsection}{0pt}{6pt}{4pt}

\defcitealias{Ricarte&Natarajan2018b}{RN18b}

\begin{document}

%\vskip 2cm

%\centerline{\LARGE \bf Where are the Intermediate Mass Black Holes?}
% * <ann.hornschemeier@nasa.gov> 2018-02-09T16:08:40.683Z:
% 
% OK, so in the last decadal it seemed they were asking for key science questions so it might be good to pose a scientific question in the title, with an implication that understanding IMBH are key to addressing it.     I would also consider leading the introduction with the mention of SMBH seeds being the key reason to care about IMBHs.
% 
% ^.

\thispagestyle{empty}
%%%%%%%%%%%%%%%%%%%%%%%%%%%%%%%%%%%%%%%%%%%%%%%%%%%%%%%%%%%%%%%%%%%%%%%
\begin{center}

{\large\bf Astro2020 Science White Paper}

\bigskip{\Large\bf Disentangling nature from nurture: tracing the origin of seed black holes }

\end{center}

\vfill

\noindent{\bf Thematic Areas:}
$$\vbox{\hsize 5truein \halign{# \hfil & #\hfil \cr
$\square$ Planetary Systems & $\square$ Star and Planet Formation \cr
$\checkbox$ Formation and Evolution of Compact Objects & $\checkbox$Cosmology and Fundamental Physics \cr
$\checkbox$  Stars and Stellar Evolution & $\square$ Resolved Stellar Populations and their Environments \cr
$\checkbox$ Galaxy Evolution & $\checkbox$ Multi-Messenger Astronomy and Astrophysics \cr
}}$$

\vfill

\begin{center}
{\it by}

Priyamvada Natarajan \& Angelo Ricarte
\\
Department of Astronomy, Yale University, New Haven, CT 06520.\\
{\tt priyamvada.natarajan@yale.edu}

\bigskip{\it in collaboration with}

Vivienne Baldassare$^{1}$, Jillian Bellovary$^{2,3}$, Peter Bender$^4$, Emanuele Berti$^5$, Nico Cappelluti$^6$, Andrea Ferrara$^{7}$, Jenny Greene$^8$, Zoltan Haiman$^{9}$, Kelly Holley-Bockelmann$^{10}$, Guido Mueller$^{11}$, Fabio Pacucci $^{12}$, David Shoemaker$^{13}$, Deirdre Shoemaker$^{14}$, Michael Tremmel$^{15,16}$, C. Meg Urry$^{1,15,16}$, Alexey Vikhlinin$^{17}$, Marta Volonteri$^{18}$

\end{center}

\setlength{\parindent}{0.25in}
\vspace{2cm}
{\noindent
{\bf Executive Summary:}
The origin and properties of black hole seeds that grow to produce the detected population of supermassive black holes are unconstrained at present. Despite the existence of several potentially feasible channels for the production of initial seeds in the high redshift universe, since even actively growing seeds are not directly observable at these epochs, discriminating between models remains challenging.  Several new observables that encapsulate information about seeding have been proposed in recent years, and these offer exciting prospects for truly unraveling the nature of black hole seeds in the coming years. One of the key challenges for this task lies in the complexity of the problem, the required disentangling of the confounding effects of accretion physics and mergers, as mergers and accretion events over cosmic time stand to erase these initial conditions. Nevertheless, some unique signatures of seeding do survive and still exist in: local scaling relations between black holes and their galaxy hosts at low-masses; in high-redshift luminosity functions of accreting black holes; and in the total number and mass functions of gravitational wave coalescence events from merging binary black holes. One of the clearest discriminants for seed models are these high redshift gravitational wave detections of mergers from space detectable in the milliHertz range. These predicted event rates offer the most direct constraints on the properties of initial black hole seeds. Improving our theoretical understanding of black hole dynamics and accretion will also be pivotal in constraining seeding models in combination with the wide range of multi-messenger data.
}

\clearpage
%=============================================
\setcounter{page}{1}

% -------- Introduction ---------------------
\centerline{\bf Introduction: the need for understanding seed formation}

\begin{figure}[htb!]
\centering
\includegraphics[width=0.2\textwidth]{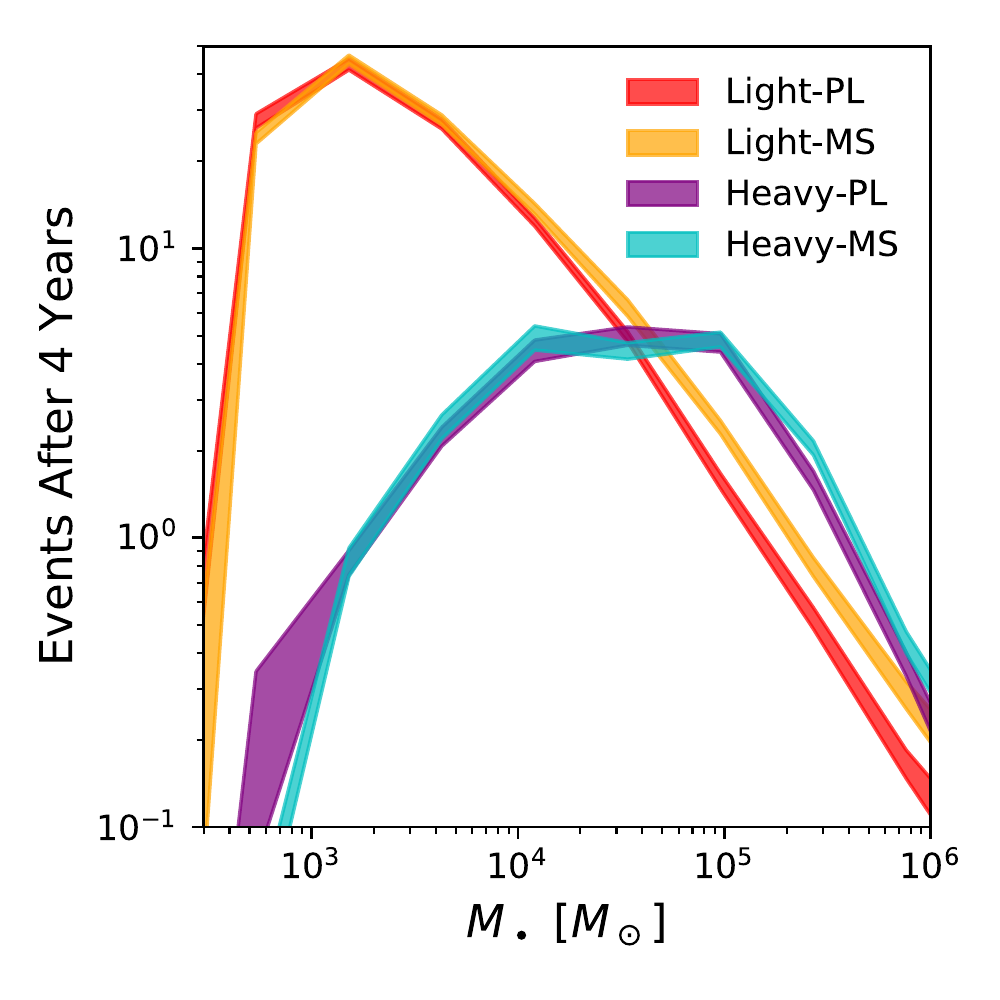}
\includegraphics[width=0.2\textwidth]{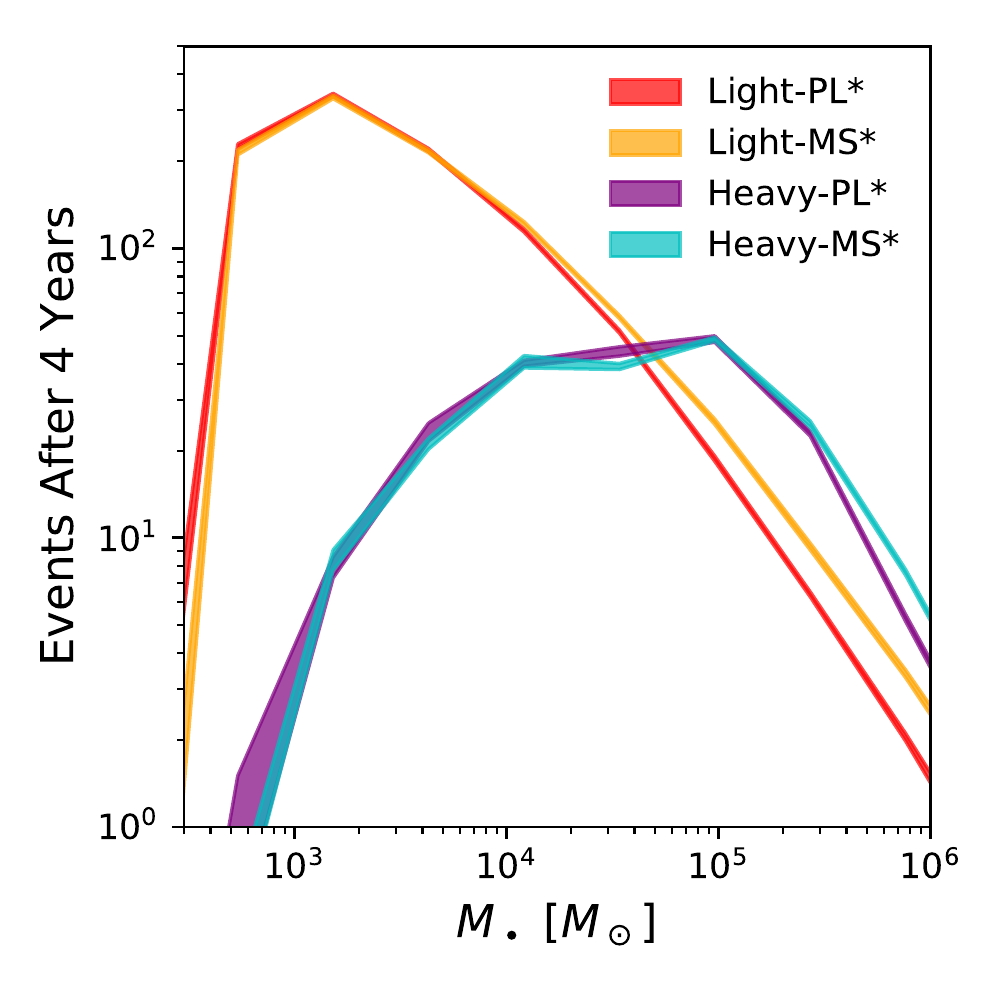}
\includegraphics[width=0.2\textwidth]{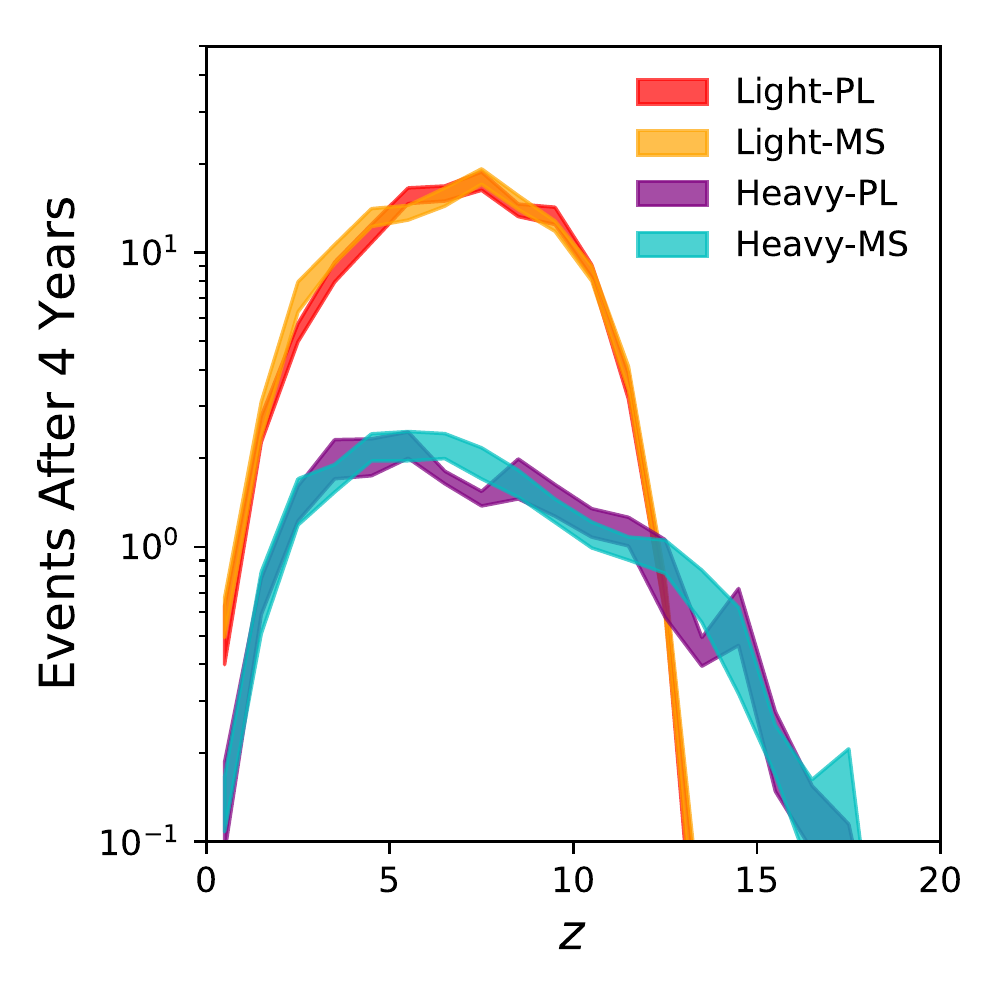}
\includegraphics[width=0.2\textwidth]{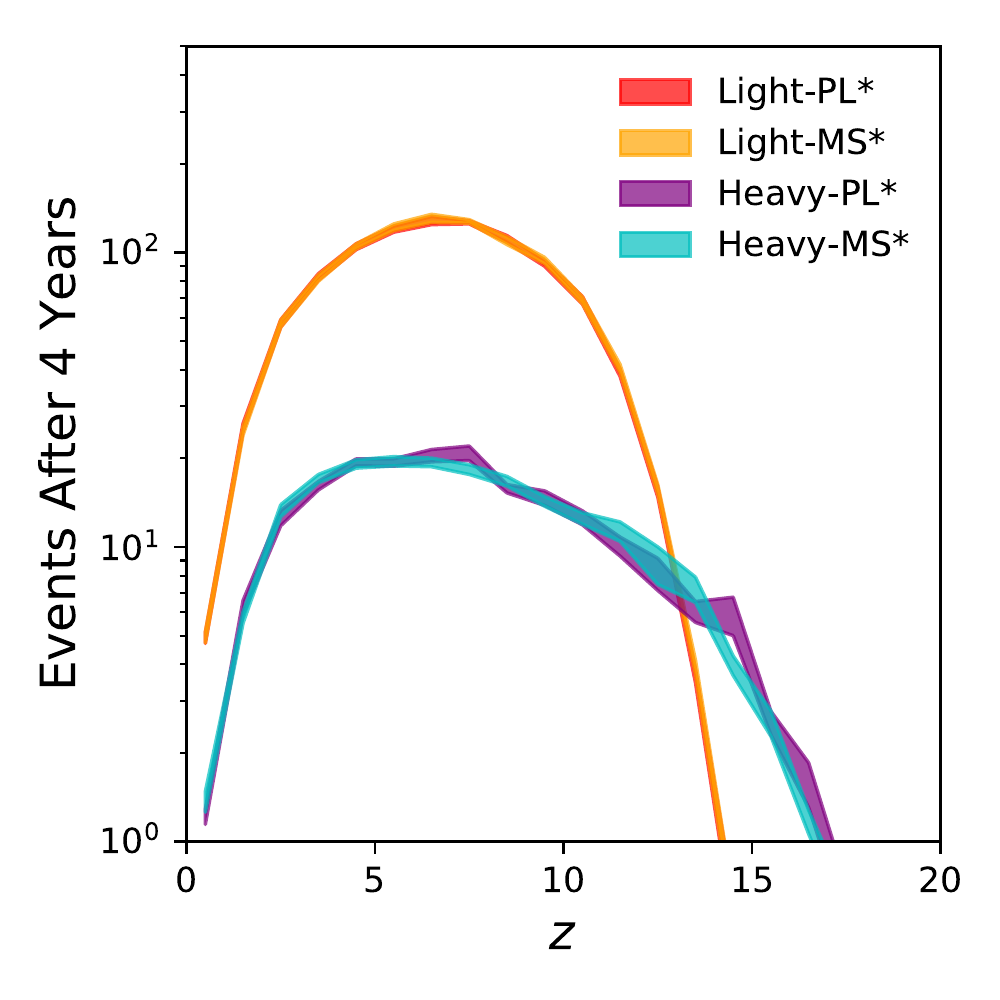}
\vspace{-3mm}
\caption{\small The chirp mass function of detectable gravitational wave events  for a 4-year  milli-Hertz GW mission like LISA as a function of chirp mass and redshift for pessimistic merger rates ({\it left}) and optimistic merger rates ({\it right}). ``Light'' and ``Heavy'' seeding scenarios are plotted, with ``MS'' and ``PL'' denoting two different plausible accretion models (Figure reproduced from \citetalias{Ricarte&Natarajan2018b}.)}
\label{fig:1}
\end{figure}

Demography of local galaxies suggests that a supermassive black hole (SMBH) is harbored in almost all galactic centers. Dynamically determined masses of these central black holes appear to correlate with several key properties of their host bulges: masses, velocity dispersion and luminosities \citep{Ferrarese&Merritt2000,Tremaine+2002,Kormendy&Ho2013}. BHs are expected to grow via mergers and accretion, but how the very first black hole seeds formed is poorly understood and is unconstrained by current observations. Quasars, powered by accretion onto SMBHs are now detected to the earliest epochs ($z\sim6-7$), when the universe was only a Gyr old. The masses for these behemoth SMBHs inferred using broad-line spectroscopy are of the order of $\sim 10^9 \ M_\odot$ at $z\sim6$, which places some constraints on a combination of the seeding and accretion conditions that produced these objects \citep{Fan+2003,Mortlock+2011,Wu+2015,Banados+2018}.  These masses require massive initial seeds and/or special growth conditions that permit extremely efficient and rapid accretion. The key issue for rapid growth is the time crunch, which requires exquisite convergence of optimal growth conditions like the presence of copious gas reservoirs to enable rapid accretion and simultaneous suppression of feedback that could disrupt inflowing gas \citep[e.g.,][]{Haiman&Loeb2001}. These conditions are more likely to be available at these earliest epochs, yet they are difficult to sustain continuously to grow seed BHs with masses of 10 - 100 $M_\odot$ to the final SMBH masses powering $z > 6$ quasars \citep{Park+2016,Pacucci+2017, Pacucci_2018_M-sigma}. This has motivated the exploration of alternate seeding prescriptions, particularly those that might produce more massive initial seeds or those in which rapid amplification of accreted mass can be achieved by finely tuning cosmic conditions \citep[see][for reviews]{Volonteri&Bellovary2012,Haiman2013,Natarajan2014, Woods_2018}. Related aspects of seeding are discussed in white papers by Haiman et al. (2019), Bellovary et al. (2019), and Pacucci et al. (2019).
\vspace{-0.4cm}
\flushleft{\bf Seed Formation Models: Light and Massive initial seeds} SMBH seeds are thought to have formed out of pristine gas in the very early universe ($z\gtrsim 15$).  With atomic hydrogen as the only coolant, abnormally large Jeans masses can be obtained. Early studies of the formation of the first stars indicated that the initial mass function was tilted high  compared to today \citep{Bromm+2002}. In more recent state-of-the-art simulations, clouds fragment more readily, resulting in lower mass stars than previously found \citep{Clark+2011,Greif+2012,Latif+2013,Hirano+2014,Stacy+2016}.  In any case, the remnants from this first generation of stars yield early light black hole seeds with masses on the order of 100 $M_\odot$. While forming light initial seeds from Pop III remnants might almost seem fairly natural and inevitable in the context of current structure formation models, growing them to $\sim\,10^9\,{M_{\odot}}$ in less than a Gyr is still prohibitive. Light seeds have been shown to accrete gas with lower duty cycles and at sub-Eddington rates due to the typical cosmic environments that might harbor them \citep{Inayoshi+2016,Pacucci+2017}.  Heavy seeds, produced from the direct collapse of pre-galactic gas disks in pristine halos were therefore proposed as an alternative path to account for the SMBHs powering the brightest, highest redshift quasars. In this Direct Collapse Black Hole (DCBH) seed formation picture, gas in early disks that go dynamically unstable and rapidly siphon mass to the center due to active and effective angular momentum dissipation via non-axisymmetric structures like bars on small scales, would grow massive central objects \citep{Oh&Haiman2002,Bromm&Loeb2003,Lodato&Natarajan2006,Begelman+2006}.  DCBHs form when fragmentation of these gas disks is prevented by suppressing cooling by dissociating any molecular hydrogen that might form \citep{Shang+2010,Visbal+2014,Regan+2017}.\citet{Lodato&Natarajan2006,Lodato&Natarajan2007} first demonstrated that connecting the larger scale cosmological context with the fate of these pristine gas disks could be used to derive the initial mass function for heavy seeds. Today, radiative and hydrodynamical cosmological simulations are available to probe the availability of direct collapse seeding sites. These studies agree that seeding sites are common enough in the early universe to explain the existence of observed ($z\sim 6-7$) quasars, but are likely too rare to account for the existence of the typical SMBH in a local, $L_*$ galaxy \citep{Habouzit+2016,Wise+2019}. In addition to direct collapse of gas, heavy seeds can also result when an initially light seed under the right set of circumstances undergoes extremely rapid growth via super-Eddington accretion especially when aided by mergers with other seeds \citep{Alexander&Natarajan2014,Dunn+2018} and  \citep{Volonteri&Rees2005}.  Such rapidy amplified (in ~$10^6$ yr). Intermediate mass black hole seeds that result from the dynamical core collapse of the first star clusters could also be a channel for early seed formation \citep{Omukai+2008,Devecchi&Volonteri2009,Stone+2017}. 
\vspace{-0.25cm}
\flushleft{\bf Mapping seeding to observables} Semi-analytic/empirical models are currently the most efficient tool for predicting the signatures of SMBH seeding that persist until observable redshifts \citep[e.g.,][]{Volonteri+2003,Somerville+2008,Tanaka&Haiman2009}.  These models deploy results of detailed cosmological simulations into simpler analytic models with parameters to quickly make predictions that are representative of large volumes and low-mass halos, that are computationally expensive for cosmological simulations to resolve. Improved and sophisticated current models formulated in the same vein offer novel ways of probing seed masses and have helped derive new sets of observables that will help to powerfully discriminate between the light and massive seeding channels \citep{Ricarte&Natarajan2018a,Ricarte&Natarajan2018b}. The latter model will henceforth be referred to as \citetalias{Ricarte&Natarajan2018b}. Seeding channels have two salient properties:  their initial masses and their abundances. In state-of-the-art models an initial population of seeds is generated with simple prescriptions based on a merger-triggering paradigm.  Light seeds are randomly drawn from an IMF ranging from 30-100 solar masses representing PopIII remnant masses, while heavy seeds are assigned according to the dynamical model of \citet{Lodato&Natarajan2006}.  Within the larger structure formation paradigm, light seeds are populated in dark matter halos above a $3.5 \sigma$ peak, but heavy seeds are only placed more rarely in halos with a narrow range of velocity dispersions -- those which are massive enough to allow atomic cooling, but not too massive to cause fragmentation and star formation.  Compared to recent hydrodynamical simulations, such models allow for an optimistically high number density of DCBHs to explore seeding signatures more cleanly. Galaxy mergers are believed to be the drivers of SMBH growth as these also trigger accretion episodes.  In this picture, when a major merger occurs, SMBHs accrete at the Eddington limit with a duty cycle of unity until they reach the local observed $M_\bullet-\sigma$ relation. Since it is currently thought that not all AGN are triggered by mergers, a long-lived sub-dominant steady accretion mode is also modeled. With the modeling flexibility that permits incorporating multi-wavelength observational constraints more readily, in one model - the AGN Main Sequence (MS) model - the black hole accretion rate for instance, is set to a thousandth of the star formation rate, as inferred from stacked observations of star forming galaxies \citep{Mullaney+2012}. Another steady mode explored is the Power Law (PL) model, in which black holes instead accrete at a rate drawn from a universal power-law Eddington ratio distribution tuned to roughly reproduce the observed luminosity functions. The potency of this modeling approach derives from the power to simultaneously explore seeding and accretion physics that eventually helps in disentangling these two confounding effects.
   %% ===== SECTION: observational challenges ==============
\centerline{\bf Discriminating seed formation channels: observational signatures and challenges}\\
Growth via accretion and mergers works to erase the initial seeding conditions for black holes, so the goal is to search for distinct signatures that survive, both at high and low-redshift.
\vspace{-0.25cm}
\flushleft{\bf Signatures of seeding in local relations} While one might not expect signatures from seeding at $z\gtrsim 15$ to persist at $z=0$, modeling suggests that they may be found in low-mass galaxies where central SMBHs have hopefully not evolved too much from their initial conditions \citep{Bellovary+2019,Micic+2011}.  Earlier work, claimed that heavy seeds exhibit larger scatter at the low-mass end of the $M_\bullet-\sigma$ relation than light seeds \citep{Volonteri&Natarajan2009}. Yet in \citetalias{Ricarte&Natarajan2018b} it is found that the scatter at the low-mass end of the $M_\bullet-\sigma$ relation is determined primarily by uncertain accretion prescriptions rather than seed mass. While the accretion prescriptions led to different outcomes, Light and Heavy seeds are not surprisingly similar.  In theory, as originally proposed by \cite{Volonteri+2008} the black hole occupation fraction was a proposed observable discriminant, unfortunately, this quantity is never directly measured. A new calibrated version of the occupation fraction, proposed in \citetalias{Ricarte&Natarajan2018b} could sift out the effect of accretion models from seeding, but such inferences are subject to large uncertainties regarding SMBH growth in low-mass hosts. Interestingly, many state-of-the-art cosmological simulations characterizing SMBH growth in low-mass hosts find that their growth is stunted \citep{Sanchez+2018}.  In dwarf galaxies, the BHs wander away from the center of the gravitational potential and may find their accretion regulated not only by their own accretion and feedback behavior, but also by feedback from supernovae \citep{Dubois+2015,Habouzit+2017,Angles-Alcazar+2017,Bellovary+2019}.
\vspace{-0.25cm}
\flushleft{\bf Signatures of seeding in high-redshift luminosity functions} Luminosity functions also contain information about a combination of seeding and accretion. The computed bolometric luminosity functions of AGN out to $z=12$ reveal that the light and heavy seed models do diverge. These high redshift luminosity functions will be easily accessible with the next-generation X-ray telescopes. Therefore, despite model degeneracies, measurement of these high-redshift X-ray luminosity functions which are well within reach observationally with proposed future facilities will allow us to constrain SMBH growth in a multi-dimensional parameter space, including seed masses, occupation fractions, and Eddington ratio distributions. Significant advances are currently being made via hydrodynamical cosmological simulations to guide accretion models, to help with separating out initial seeding effects. The discrimination between seeding models shows up primarily at high redshift and therefore data from these epochs will prove to be pivotal.
\vspace{-0.25cm}
\flushleft{\bf Signatures of seeding in gravitational wave events} Models currently find that the cleanest discriminant of initial seeding derive from observations of resolved gravitational wave events that measure the rate of black hole mergers irrespective of our assumptions about accretion \citep{Sesana+2005,Sesana+2011,Klein+2016}. Merging black holes produce gravitational waves, and stellar-mass black hole mergers have now been detected by the Laser Interferometer Gravitational-Wave Observatory (LIGO) \citep[e.g.,][]{LIGO2016}.  For higher mass black hole mergers, a lower frequency milliiHertz space interferometer will be the instrument needed to probe the initial seed population \citep{LISA2017}.  The planned LISA mission has the sensitivity to detect the high-redshift mergers in low-mass hosts, that encapsulate information on the initial masses and occupation fractions of seeding. Current model predictions for the event rate and dependence of such events on the mass ratios of merging black holes, and their redshifts robustly indicate that seed masses can be inferred. A set of predictions adopting the signal-to-noise calculation outlined in \citet{Sesana+2005,Sesana+2007} are presented in Fig.~\ref{fig:1} (details can be found in \citetalias{Ricarte&Natarajan2018b}). Unlike the other observables discussed thus far, estimates of the merging black hole population do not depend on their instantaneous accretion state. However, they do depend on the dynamics regarding the formation and shrinking of black hole binaries which are poorly constrained from kpc \citep{Tremmel+2017,Pfister+2019} to pc \citep[e.g.,][]{Colpi+2014} scales, leading to a large spread in detection rates ranging from $\sim 1$--$100$ events during a putative 4-year mission depending on the assumptions made \citep{Sesana+2011,Klein+2016}.  Light seed models produce significantly more gravitational wave events than heavy seeds at low chirp masses. This is a direct reflection of the initial seed masses and occupation fractions.  After a 4-year mission, with conservative assumptions on the dynamics, $\sim 20$ detected mergers for heavy-seeds, and $\sim 100$ detected mergers for a light-seeding scenario are predicted.  Under the more optimistic assumption that no mergers stall dynamically, the predicted detection rates increase by an order of magnitude, both these cases are illustrated in Fig.~1 \citep{Khan+2013}.

%%%%%   figure   %%%%%%%

    %% ===== SECTION: Summary and Closing==============
\vspace{-0.5cm}    
 \centerline{\bf Summary}
 
The origin of seed black holes is fiercely debated and is one of the key open questions in cosmology today. The cosmic assembly history of black holes is a result of the combination of seeding, accretion physics and dynamical processes and disentangling these is a challenging task. Although the cumulative accretion history over cosmic time tends to erase initial seeding conditions, some unique signatures persist that can help distinguish between light and heavy seeds. Current models that track SMBH assembly over cosmic time show that a powerful set of new observables from multi-wavelength data sets can help distinguish seeding scenarios. Key observables are the event rates of SMBH binary mergers. These coalescence rates as a function of redshift and chirp mass offer unique discrimination between initial black hole seeding models. A 3 spacecraft gravitational wave detector, such as the proposed LISA mission, has the sensitivity and power for these critical observations that offer the prospect of profound insights into the physics of seeding. The combination of gravitational wave events and multi-messenger electromagnetic probes will be critical to obtain a self-consistent and comprehensive view of black hole formation and assembly.   
\newpage
\bibliographystyle{aasjournal}

\centerline{\bf Instituitional Affliations of Authors}

\noindent $^1${Department of Astronomy, Yale University, New Haven, CT, USA}\\
$^2${City University of New York, Queensborough Community College, NY,  USA}\\
$^3${American Museum of Natural History, New York, NY, USA}\\
$^4${JILA, University of Boulder, Colorado, CO, USA}\\
$^5${Department of Astronomy and Joint Space-Science Institute, The University of Maryland, USA }\\
$^6${University of Miami, Miami, Florida, USA}\\
$^7${SNS, Pisa, Italy}\\
$^8${Department of Astrophysical Sciences, Princeton University, Princeton, NJ, USA}\\
$^9${Department of Astronomy, Columbia University, New York, NY, USA}\\
$^{10}${Vanderbilt University, Nashville, TN, USA}\\
$^{11}${University of Florida, Gainesville, FL, USA}\\
$^{12}${Kapteyn Astronomical Institute, University of Groningen, Groningen, The Netherlands}\\
$^{13}${Department of Physics, Yale University, New Haven, CT, USA}\\
$^{14}${Yale Center for Astronomy and Astrophysics, New Haven, CT 06520, USA}\\
$^{15}${EAPS, M.I.T., Cambridge, MA, USA}\\
$^{16}${Georgia Tech, Atlanta, USA}\\
$^{17}${Harvard-Smithsonian Center for Astrophysics, Cambridge, MA, USA}\\
$^{18}${IAP, Paris, France}

\bibliography{newseeds_final}

\end{document}